\documentclass[]{aa} 
\usepackage{graphics} 
\begin{document} 

\thesaurus{4(02.16.2; 11.09.1 NGC 3627; 11.13.2; 11.19.2; 13.18.1)} 
\title{Magnetic fields and strong density waves in the interacting 
galaxy NGC~3627} 
\author{M. Soida\inst{1} 
\and M. Urbanik\inst{1} 
\and R. Beck\inst{2} 
\and R. Wielebinski\inst{2}} 
\institute{Astronomical Observatory, Jagiellonian University, Krak\'ow, 
Poland 
\and 
Max-Planck-Institut f\"ur Radioastronomie, Postfach 2024, D-53010 Bonn, 
Germany 
} 
\offprints{M. Urbanik} 
\mail{urb@oa.uj.edu.pl} 

\date{Received 28 January 1999 / Accepted 8 February 1999 } 

\maketitle 
\titlerunning{Magnetic field and strong density waves ...} 
\authorrunning{M. Soida} 

\begin{abstract} 

We present 10.55~GHz observations of the strong-density-wave spiral 
NGC~3627. Total power and polarization maps with a resolution of 1\farcm 
13, sensitive to a weak extended emission were obtained. In the analysis 
we used also available data in the CO and H$\alpha$ lines. 

The total power brightness distribution shows two equal\-ly bright 
sources close to the bar ends, coincident with similar peaks in CO and 
H$\alpha$. The strong central peak of the CO emission does not 
correspond to a detectable central source in radio continuum. A weak 
H$\alpha$ flux from this region is thus due not only to a strong 
absorption but may also indicate a low star formation level in the 
central molecular complex. The equally strong total power peaks at the 
bar ends do not reflect the asymmetry of the H$\alpha$ emission, the 
latter being stronger at the northern bar end. The H$\alpha$ asymmetry 
is likely to be due to differences in absorption. 

The polarized emission has the form of two asymmetric lobes with 
B-vectors running parallel to the optical arms. The stronger lobe is 
located at the position of the dust lane in the western arm while the 
weaker one falls on the middle of the interarm space in the NE disk. 
Smooth polarized emission away from any spiral structures was also 
detected. Despite the strong density waves, many polarization properties 
of this galaxy like the large-scale distribution of polarized intensity 
or azimuthal variations of magnetic pitch angles can be reasonably 
explained by the presence of an axisymmetric, dynamo-type magnetic field 
component. However, extra effects like the depolarization of the 
southern segment of the eastern arm by vertical fields above 
star-forming regions, as well as some compressional enhancement of 
regular fields in the western arm seem necessary to explain our 
polarization data. 

\end{abstract} 

\keywords{ Polari\-zation -- Gala\-xies:indivi\-dual:NGC~3627 -- 
Gala\-xies:magne\-tic fields -- Gala\-xies:spiral -- 
Radio conti\-nuum:gala\-xies } 

\section{Introduction} 

The role of gas flows in spiral arms upon the galactic magnetic field 
evolution is a lively debated issue. Theories of the field amplification 
by small scale turbulent motions (e.g. axisymmetric dynamo, Wielebinski 
\&~Krause 1993), which well reproduce the polarization properties of 
galaxies with weak density waves (Urbanik et~al. 1997), do not need any 
spiral arm flows. In other theories (e.g. Chiba 1993), the density wave 
perturbations are the main agent amplifying the galactic magnetic field. 
The importance of the latter process compared to the dynamo action may 
be a function of the density wave strength. Though successful attempts 
to model the magnetic field evolution driven by the dynamo and spiral 
arms or bars have been made (Mestel \&~Subramanian 1991, Subramanian 
\&~Mestel 1993, Moss 1997, Moss et~al. 1998), this issue has still very 
poor observational grounds. 

Existing observations of large, nearby spirals (Beck et~al. 1996) do not 
allow to state whether in case of strong density waves the magnetic 
field evolution becomes dominated by processes in spiral arms. Strong 
density wave signatures are present in M51 and in the inner disk of M81. 
In M51 the polarization B-vectors follow local structural details of 
dust lanes (Neininger \&~Horellou 1996), as expected for a magnetic 
field dominated by density-wave compression. However, the spiral arms 
are too tightly wound to separate this field component from a possible 
dynamo-generated one, which is usually distributed more uniformly in the 
disk. No clear magnetic field component related to density waves has 
been identified in M81, however its inner disk shows a subtle network of 
local compression regions filling the whole interarm space (Visser 
1980). 

NGC~6946, NGC~4254 and the outer parts of M81 do not show strong density 
wave signatures. Their magnetic fields form either broad ''magnetic arms`` 
in the middle of the interarm space (NGC~6946, Beck \&~Hoernes 1996) or 
smoothly fill the interarm space (M81, Krause et~al. 1989). In NGC~4254 
a coherent spiral pattern of polarization B-vectors exists even in 
regions of completely chaotic optical structures (Soida et~al. 1996). 
The nearby well-studied spirals also have enhanced star formation in 
spiral arms, destroying the regular fields. We note that the detection 
of smoothly distributed dynamo-type fields needs a very good sensitivity 
to extended polarized emission, which is not always ensured by 
high-resolution studies. 

In this paper we present observations with good sensitivity to smooth, 
extended structures to check whether a galaxy with very strong signs of 
density waves may have a global magnetic field dominated by the 
component caused by density wave action. We obtained 10.55~GHz total 
power and polarization maps of the spiral NGC~3627 interacting within 
the Leo Triplet (Haynes et~al. 1979). The galaxy has a bar and two 
spiral arms with a broad interarm space discernible even with a modest 
resolution (see Fig.~1). The western arm contains a long dust lane 
tracing large-scale gas (and possibly frozen-in field) compression, 
accompanied by little star formation (cf. H$\alpha$ map by Smith et~al. 
1994). The middle part of the arm is unusually straight, bending sharply 
in the outer disk. The eastern arm has a heavy dust lane in its southern 
part, breaking into a subtle network of filaments in its northern half. 
The southern dust lane segment is accompanied by a chain of bright 
star-forming regions. Reuter et~al. (1996) found perturbations of the 
galaxy's CO velocity field possibly due to streaming motions related to 
spiral arms. NGC~3627 has been also observed in the far infrared by 
Sievers et~al. (1994). A total power map at 1.49~GHz using the VLA 
D-array was made by Condon (1987). 

\section{Observations and data reduction} 

\begin{figure*} 
\resizebox{12cm}{!}{\includegraphics{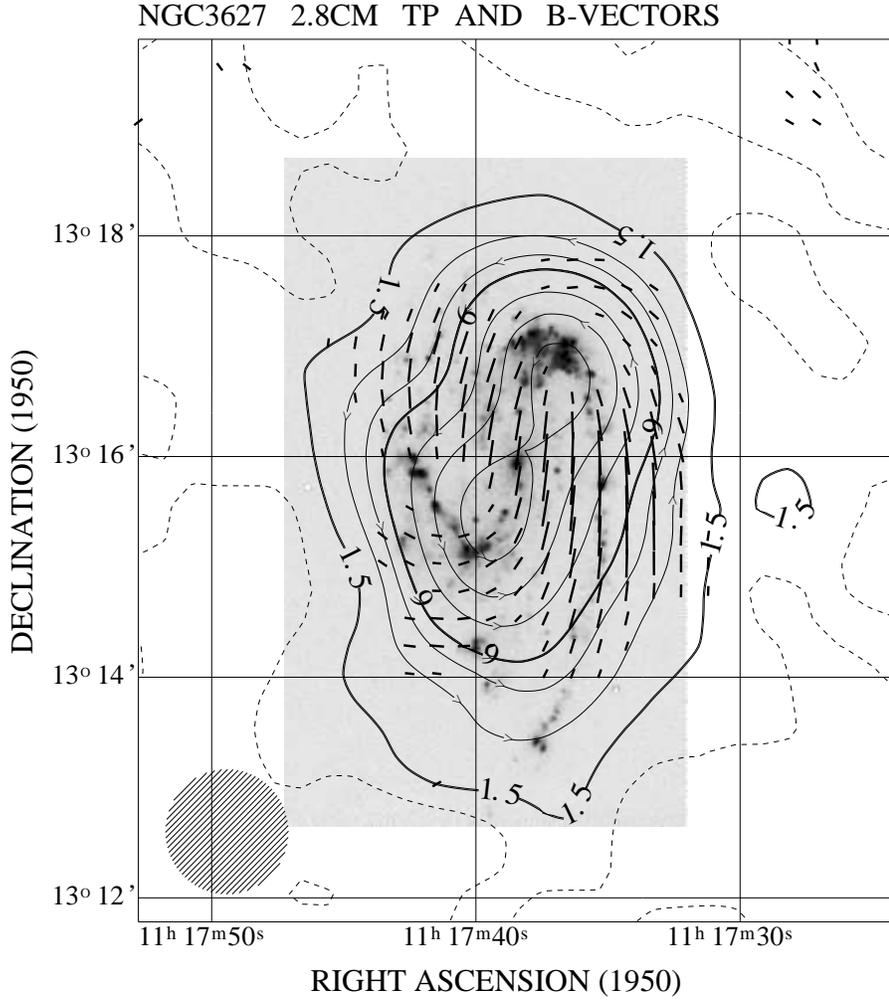}} 
\parbox[b]{55mm}{ 
\caption{ 
The total power contour map of NGC~3627 at 10.55~GHz with B-vectors of 
polarized intensity superimposed onto the H$\alpha$ image from Smith 
et~al. (1994). The resolution is 1\farcm 13. The contour levels are 1.5, 
4, 6.5~mJy/b.a., then 9, 14, 19$\ldots$ etc. mJy/b.a.. The first contour 
corresponds to about 2.5$\sigma$ r.m.s. noise. The dashed contour shows 
the zero level 
} 
\label{fig1} }
\end{figure*} 

The total power and polarization observations at 10.55 GHz were 
performed in May 1993, as well as in April and May 1994 using the 
four-horn system in the secondary focus of the Effelsberg 100-m MPIfR 
telescope (Schmidt et~al. 1993). With 300~MHz bandwidth and $\sim 40$~K 
system noise temperature, the r.m.s. noise for 1~sec integration and 
combination of all horns is $\sim 2$~mJy/beam area in total power and 
$\sim 1$~mJy/beam area in polarized intensity. 

Each horn was equipped with two total power receivers and an IF 
polarimeter resulting in 4 data channels containing the Stokes 
parameters I, Q and U. The telescope pointing was corrected by making 
cross-scans of Virgo~A at time intervals of about 2~hours. As flux 
calibrator the highly polarized source 3C286 was observed. A total power 
flux density of 4450~mJy at 10.55~GHz has been adopted using the 
formulae by Baars et~al. (1977). The same calibration factors were used 
for total power and polarized intensity, yielding a mean degree of 
polarization of 12.2\%, in reasonable agreement with other published 
values (Tabara \&~Inoue 1980). 

In total 29 coverages of NGC~3627 in the azimuth-elevation frame were 
obtained. The data reduction process was performed using the NOD2 data 
reduction package (Haslam 1974). By combining the information from 
appropriate horns, using the ''software beam-switching`` technique (Morsi 
\&~Reich 1986) followed by a restoration of total intensities (Emerson 
et~al. 1979), we obtained for each coverage the I, Q and U maps of the 
galaxy. All coverages were then combined using the spatial frequency 
weighting method (Emerson \&~Gr\"ave 1988), yielding the final maps of 
total power, polarized intensity, polarization degree and polarization 
position angles. A digital filtering process, which removes spatial 
frequencies corresponding to noisy structures smaller than the telescope 
beam, was applied to the final maps. A special CLEAN procedure to remove 
instrumental polarization was applied to the polarization data. The 
original beam of our observations was 1\farcm 13. With the 
distance modulus of $-30\fm37$ given by Ryan \&~Visvanathan (1989), 
corresponding to a distance of 11.9~Mpc, our beamwidth is equivalent to 
3.9~kpc in the sky plane. In the galaxy's disk plane this corresponds to 
3.9 and 8 kpc along major and minor axes, respectively. 

\section{Results} 

\subsection{Total power emission} 

The total power map at the original resolution with B-vectors of 
polarized intensity is shown in Fig.~1. The map has an r.m.s. noise of 
0.6~mJy/b.a. Bright total power peaks are found at the bar ends, where 
both the CO(1-0) and CO(2-1) maps by Reuter et~al. (1996) as well as the 
H$\alpha$ map by Smith et~al. (1994) show large accumulations of 
molecular gas and young star formation products. There is no indication 
of a bright central source. 

The outer disk shows a remarkable asymmetry. The total power emission is 
considerably more extended and decreases more smoothly towards the south 
than to the north. In this respect it resembles the optical, CO and 
H$\alpha$ morphology: the western arm running southwards extends to a 
considerably larger distance from the centre than does the eastern one. 

A slight extension towards the east at RA$_{1950}$ of about 
$11^{h}17^{m}45^{s}$ and Dec$_{1950}$ of about $13\degr 17\arcmin$ is 
also seen in the map by Urbanik et~al. (1985) and must be real. It has 
no optical counterpart but corresponds roughly to the region where 
Haynes et~al. (1979) found a counter-rotating HI plume, probably caused 
by tidal interactions within the Leo Triplet. 

\begin{figure} 
\resizebox{\hsize}{!}{\includegraphics{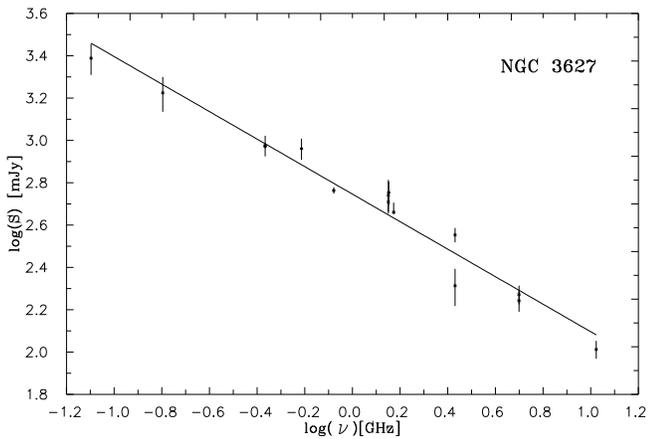}} 
\caption{ 
The integrated radio spectrum of NGC~3627 
} 
\label{fig2} 
\end{figure} 

The integration of the total power map in elliptical rings using an 
inclination of $67\fdg 5$ and a position angle of $173\degr$ (both taken 
from the Lyon-Meudon Extragalactic Database) yields an integrated flux 
density at 10.55~GHz of 103$\pm$10~mJy within the radius of 20~kpc, very 
close to the total flux obtained by Niklas et~al. (1995). This value has 
been combined with available data at lower frequencies collected in 
Table 1. All values have been converted to the flux density scale by 
Baars et~al. (1977). A weighted power law fit to the data yields a mean 
spectral index of 0.64$\pm$0.04 ($S_{\nu}\sim\nu^{-\alpha}$). As the 
deviations from a single power law are comparable to the errors in the 
observed flux densities (Fig.~2), a spectral index of 0.64 has been 
adopted for the whole frequency range between 80~MHz and 10.7~GHz. 

\begin{figure*} 
\resizebox{12cm}{!}{\includegraphics{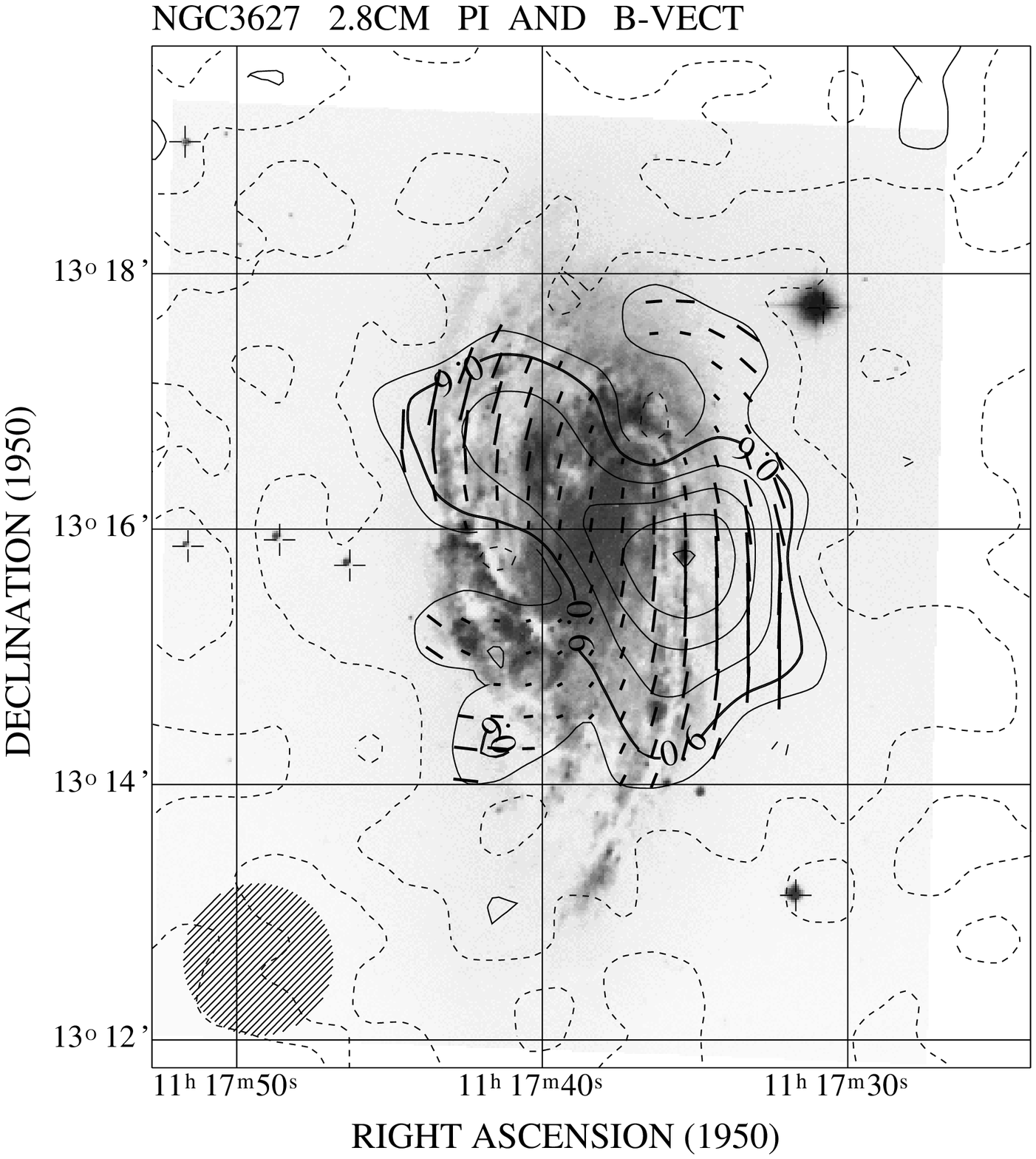}} 
\parbox[b]{55mm}{ 
\caption{ 
The contour map of the polarized intensity of NGC~3627 at 10.55~GHz with 
B-vectors of the polarization degree superimposed onto an optical image 
(Arp 1966). The resolution is 1\farcm 13. The contour levels are 
0.4~mJy/b.a., then from 0.6~mJy/b.a. with an increment of 0.4~mJy/b.a.. 
The first thin contour corresponds to about 2.2$\sigma$, and the first 
thick one to about 3.3$\sigma$ r.m.s. noise. The dashed contour shows 
the zero level 
} 
\label{fig3} }
\end{figure*} 

\begin{table}[th] 
Table 1. The integrated radio spectrum of NGC~3627 
\begin{center} 
\begin{tabular}{rrrl} \hline 
Frequ-&Flux &error & References \\ 
ency &density& & \\\ 
 [GHz] & [mJy] & [mJy] & \\ \hline 
 0.080 & 2448 &$\pm$ 410 & Slee (1972) \\ 
 0.160 & 1680 &$\pm$ 315 & Huchtmeier (1975) \\ 
 0.430 & 946 &$\pm$ 106 & Lang \& Terzian (1969) \\ 
 0.430 & 938 &$\pm$ 20 & Israel \& van der Hulst (1983) \\ 
 0.611 & 915 &$\pm$ 105 & Lang \& Terzian (1969) \\ 
 0.835 & 580 &$\pm$ 20 & Israel \& van der Hulst (1983) \\ 
 1.415 & 551 &$\pm$ 100 & de la Beaujardiere et~al. (1968) \\ 
 1.415 & 512 &$\pm$ 51 & Hummel (1980) \\ 
 1.420 & 567 &$\pm$ 72 & Whiteoak (1970) \\ 
 2.695 & 359 &$\pm$ 28 & de Jong (1967) \\ 
 2.695 & 206 &$\pm$ 41 & Kazes et~al. (1970) \\ 
 5.000 & 187 &$\pm$ 19 & Whiteoak (1970) \\ 
 5.000 & 175 &$\pm$ 20 & Sramek (1975) \\ 
 10.550 &103 &$\pm$ 10 & This paper \\ 
\hline 
\end{tabular} 
\end{center} 
\end{table} 

\subsection{Polarized intensity} 

Our pola\-rized inten\-sity map has an r.m.s. noise of 0.18 mJy/b.a.. It 
shows two asymmetric lobes with B-vectors locally parallel to the 
principal arms (Fig.~3). The strongest peak of the polarized brightness, 
with the polarization degree reaching locally 25\%, is located west of 
the galaxy's centre, at the position of the unusually straight dust lane 
segment (see also Fig.~5). No bright star-forming regions are present 
there. The second, weaker peak does not coincide with a prominent dust 
lane but is located in the interarm region between the northern segment 
of the eastern arm and the bar where only small, barely visible dust 
filaments are present. No polarization was detected in the vicinity of a 
particularly heavy dust lane segment in the southern part of the eastern 
arm at RA$_{1950}$ of about $11^{h}17^{m}42^{s}$ and Dec$_{1950}$ of 
$13\degr 15\arcmin 30\arcsec$, accompanied by a chain of star-forming 
regions. 

\begin{figure} 
\resizebox{\hsize}{!}{\includegraphics{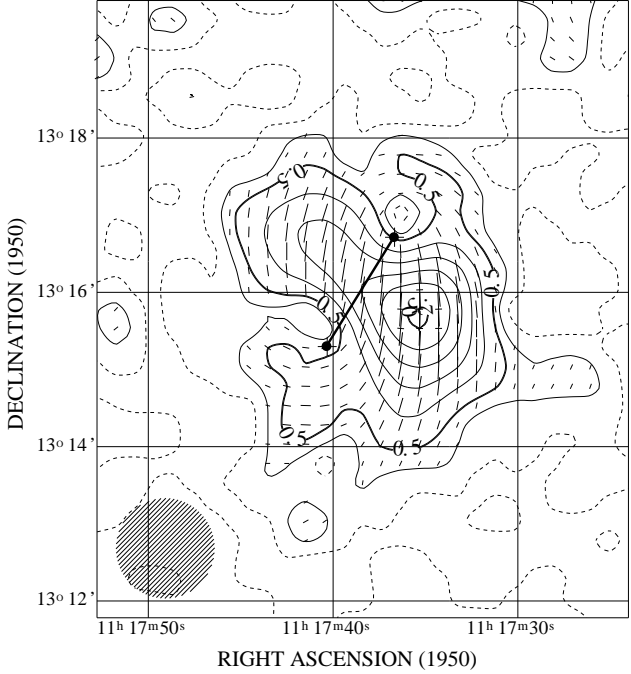}} 
\caption{ 
The contour map of the polarized intensity of NGC 3627 with B-vectors 
proportional to the same quantity, convolved to a beam of 1\farcm 3. The 
first contour is 0.3~mJy/b.a., then contours start from 0.5~mJy/b.a. and 
increase by 0.4 mJy/b.a. The first thick contour corresponds to about 
3$\sigma$ r.m.s. noise. The dashed contour denotes the zero level. The 
thick line terminated by dots marks the position of the bar as defined 
by the CO(1-0) maxima at its ends 
} 
\label{fig4} 
\end{figure} 

With a polarization degree less than 3\% the bar ends are generally 
weakly polarized. However, clear polarization patches in the vicinity of 
RA$_{1950}$ of $11^{h}17^{m}41\fs5$ Dec$_{1950}$ of $13\degr 14\arcmin 
30\arcsec$ and RA$_{1950}$ of $11^{h} 17^{m} 35\fs3$ Dec$_{1950}$ of 
$13\degr 17\arcmin 32\arcsec$, surrounding the bar ends and being 
marginally significant in Fig.~3, exceed the $3\sigma$ noise level after 
convolving the data to a beamwidth of 1\farcm 3 (Fig.~4). The degree of 
polarization in these regions is about 5--6\%. 

The orientations of the polarization B-vectors cor\-rected to face-on 
position are shown in Fig.~5. West of the centre they run parallel to a 
straight segment of the dust lane, following its bend in the southern 
disk. In the polarized peak in the NE disk the B-vectors in the interarm 
region follow the direction of the dust lane which itself is only weakly 
polarized. 

In the above mentioned weak polarization patches near the bar ends, best 
visible in Fig.~4, the B-vectors tend to turn smoothly around the 
terminal points of the bar. No ordered optical or H$\alpha$ structures 
are present there. Close to the northern bar end the vector orientations 
smoothly join these in the western arm (see also Fig.~5). Near the 
southern bar end the B-vectors deviate strongly to the east with a large 
pitch angle. Across the unpolarized region in the southern part of the 
eastern arm their orientations jump by about $90\degr$. The question of 
possible geometrical depolarization at this position is discussed in 
detail in Sect.~4. 

\begin{figure*} 
\resizebox{12cm}{!}{\includegraphics{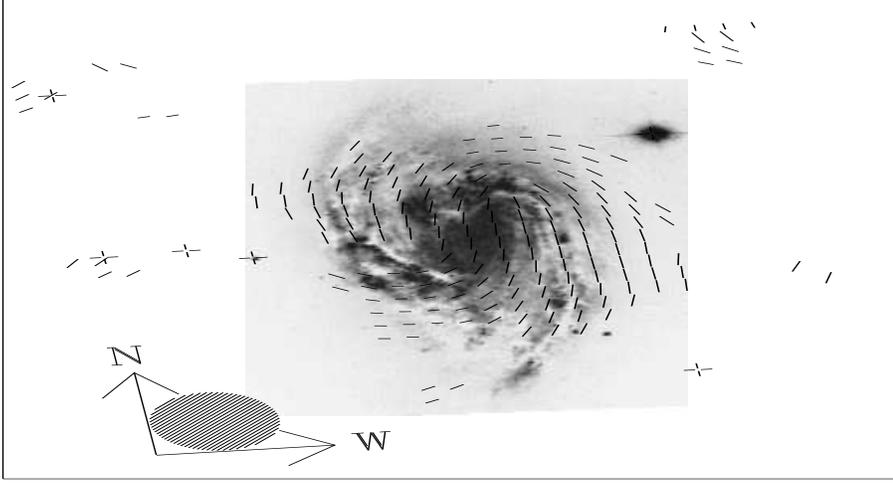}} 
\parbox[b]{55mm}{ 
\caption{ 
Overlay of B-vector directions computed from the polarization map with 
the original resolution of 1\farcm 13 onto an optical image of NGC~3627, 
both rectified to face-on position. The galaxy's major axis runs 
vertically, directions to north and west are shown by arrows 
} 
\label{fig5} }
\end{figure*} 

The integration of the polarized intensity map shown in Fig.~3 in the 
same rings as described in Sect~3.1 yields an integrated polarized flux 
density of $6.0\pm 1.8$~mJy. This implies a mean polarization degree of 
$5.8\pm 1.8$\%. An application of the formula of Segalovitz et~al. 
(1976) yields a mean ratio of regular to total field strengths $B_u/B_t 
= 0.22\pm 0.04$. 

\section{Discussion} 

\subsection{Total magnetic field strength and distribution of total 
power brightness} 

\begin{figure} 
\resizebox{\hsize}{!}{\includegraphics{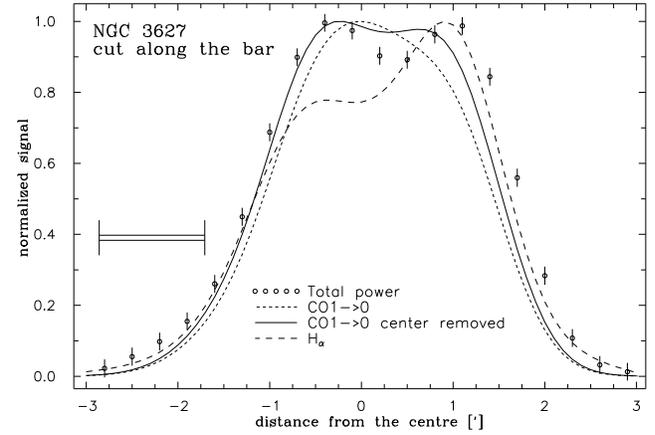}} 
\caption{ 
Cross-section along the bar of NGC~3627 in the total power brightness 
(circles and error bars) compared to that in CO(1-0) (Reuter at~al. 
1996, short-dashed line) and H$\alpha$ (Smith et~al. 1994, long dashed 
line), convolved to a common beam of 1\farcm 13 and each normalized to 
its maximum value. The CO(1-0) profile with the central peak subtracted 
from the original map is shown, too (solid line) 
} 
\label{fig6} 
\end{figure} 

The integrated radio spectrum of NGC~3627 does not show obvious 
deviations from a single power-law with a slope $\alpha=0.64$ (see 
Fig.~2). Using the integrated flux density, $\alpha=0.64$ and assuming 
the minimum-energy or energy equipartition condition we derive a mean 
total magnetic field strength of $13\pm 4\ \mu$G. The regular field 
component (assuming that it is entirely parallel to the disk) derived 
from the polarized intensity equals $3.5\pm 1.3\ \mu$G. This value 
refers to a magnetic field which is regular over scales larger than our 
beam of 4~kpc. In computing the above values we assumed a lower limit of 
the cosmic-ray spectrum of 300~MeV (cf. Beck 1991), a ratio of 
proton-to-electron density ratio of 100 (Pacholczyk 1970), and a 
nonthermal disk scaleheight of 1~kpc. The error in the total magnetic 
field strength includes an uncertainty by a factor of 2 of the 
proton-to-electron ratio, the disk thickness and the lower energy 
cutoff, as well as an unknown thermal fraction between 0\% and 40\%. The 
mean total magnetic field of NGC~3627 is stronger than average for 
spiral galaxies (Beck et~al. 1996), in spite of the low neutral gas 
content (Young et~al. 1983, Zhang et~al. 1993, see also Urbanik 1987). 

The bright total power sources at the ends of the bar lie at the 
positions of huge star-forming molecular complexes, also coincident with 
HI peaks (Zhang et~al. 1993). Their spectral index between 1.49~GHz and 
10.55~GHz derived using Condon's (1987) map is 0.73--0.75, thus 
nonthermal emission is dominating. Fig.~6 shows the cross-sections along 
the bar of the total power intensity at 10.55~GHz as well as of the 
H$\alpha$ and CO(1-0) line emission (Smith et~al. 1994, Reuter et~al. 
1996), both convolved to our resolution. All profiles were normalized to 
their maximum values. The CO profile has a peak at the galaxy's centre 
due to the emission from the central molecular complex which in the 
original maps (Reuter et~al. 1996) has the same peak brightness and 
extent as the CO complexes at the bar ends. However, the H$\alpha$ 
emission lacks the central peak being very weak in the nuclear region. 
An intense star formation in the central molecular complex with its 
manifestation in the H$\alpha$ line obscured by the dust (abundant in 
nuclear regions of spiral galaxies) is unlikely as the 10.55~GHz profile 
(Fig.~6) has a central depression, too. The total power minimum is even 
deeper than in CO(1-0) after subtracting completely the nuclear region 
from the original CO map of Reuter et~al. (1996). Thus the central 
molecular complex apparently forms stars at a much lower rate than 
aggregates of the cold gas at the bar ends. We also note that the HI map 
of Zhang et~al. (1993) shows a central depression, too. 

The H$\alpha$ emission from the southern bar end is considerably weaker 
than from the northern one. No such asymmetry exists in radio continuum. 
The CO(1-0) brightness (thus also the content of an opaque cold gas) is 
however somewhat higher at the southern than at the northern end of the 
bar. The mentioned asymmetry of the H$\alpha$ emission may thus be 
caused by a higher absorption in the southern bar end. 

The determination of the radial scale length $r_0$ of the nonthermal 
disk by fitting a beam-smoothed exponential model encounters severe 
problems because of the high inclination and emission excess at the bar 
ends. Nevertheless reasonable values are in the range 1\farcm 2--1\farcm 
8 corresponding to 4.2--6.2~kpc at the distance of 11.9~Mpc. 

\subsection{The magnetic field structure } 

\begin{figure} 
\resizebox{\hsize}{!}{\includegraphics{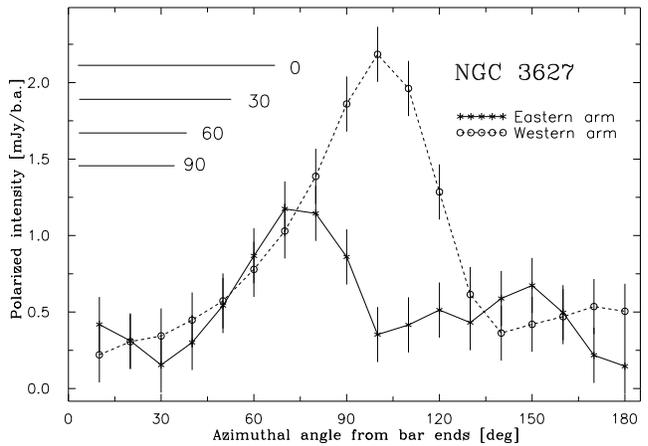}} 
\caption{ 
A comparison of changes of the polarized brightness with the azimuth 
distance from the corresponding bar ends for both polarized lobes. The 
azimuth runs anticlockwise from either the northern bar end (solid line, 
eastern lobe) or from the southern end (dotted line, western lobe). The 
polarized intensity was integrated along a ring 24\arcsec wide with a 
face-on radius of 2\arcmin, having the same inclination and position 
angle as NGC~3627, divided into sectors with an azimuthal width of 
$10\degr$. Horizontal bars show the range of azimuthal angles 
corresponding to the beam size at various azimuthal distances from the 
bar end (labelled in degrees) 
} 
\label{fig7} 
\end{figure} 

\begin{figure} 
\resizebox{\hsize}{!}{\includegraphics{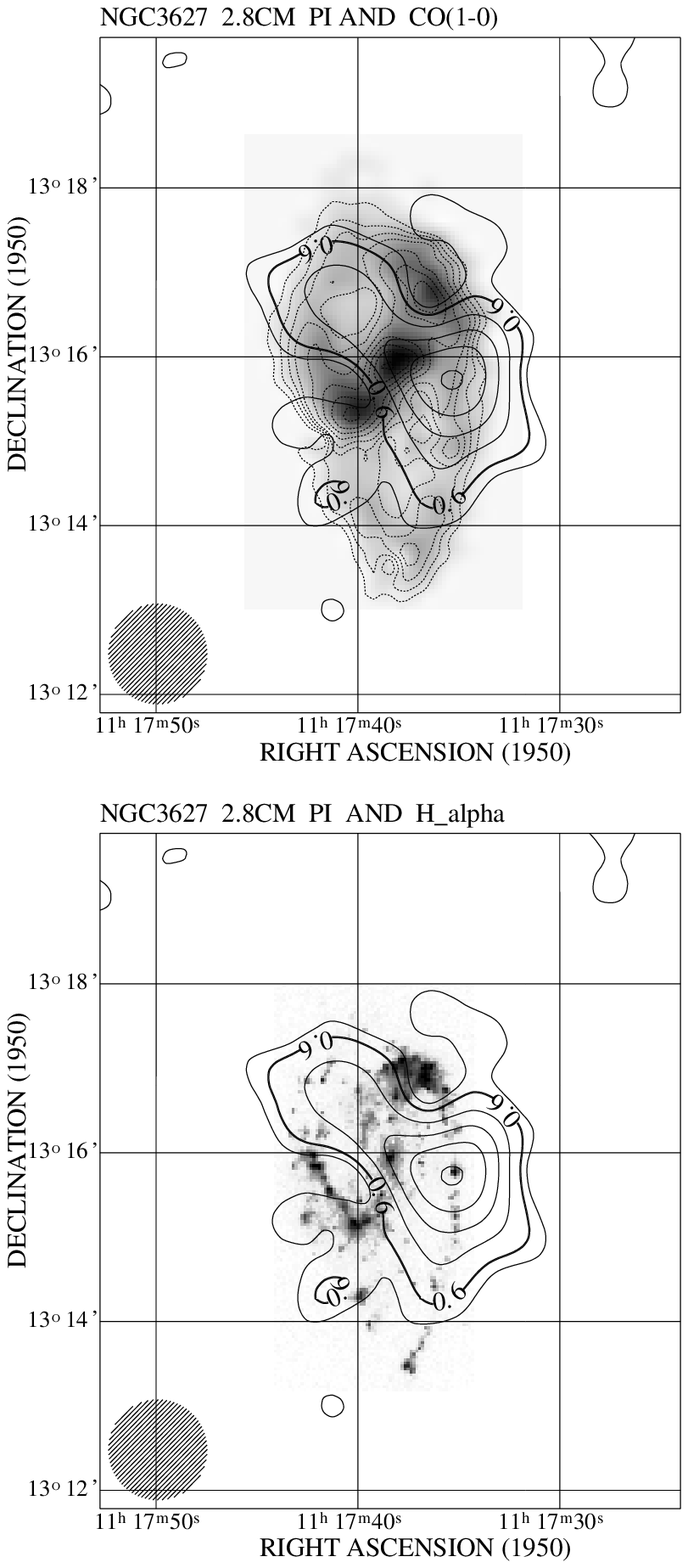}} 
\caption{ 
The contours of polarized intensity for NGC~3627 at 10.55~GHz with the 
original resolution of 1\farcm 13 overlaid onto a)~the greyscale map of 
CO(1-0) of Reuter et~al. (1996), and b)~the H$\alpha$ image of Smith 
et~al. (1994). The polarized intensity contours are the same as in 
Fig.~3 
} 
\label{fig8} 
\end{figure} 

The polarized brightness is strongly peaked at the middle of a straight 
portion of the dust lane in the western arm. To check whether the 
emission is resolved we tried to subtract the beam-smoothed point source 
at the position of the observed brightness maximum. We found that the 
polarized peak can be decomposed into an unresolved source with a 
polarized flux density of about 1.8~mJy and an extension along the 
southern part of the dust lane with a maximum polarized brightness of 
about 0.8~mJy/b.a.. The eastern lobe is however rather poorly resolved. 

\begin{figure*} 
\resizebox{12cm}{!}{\includegraphics{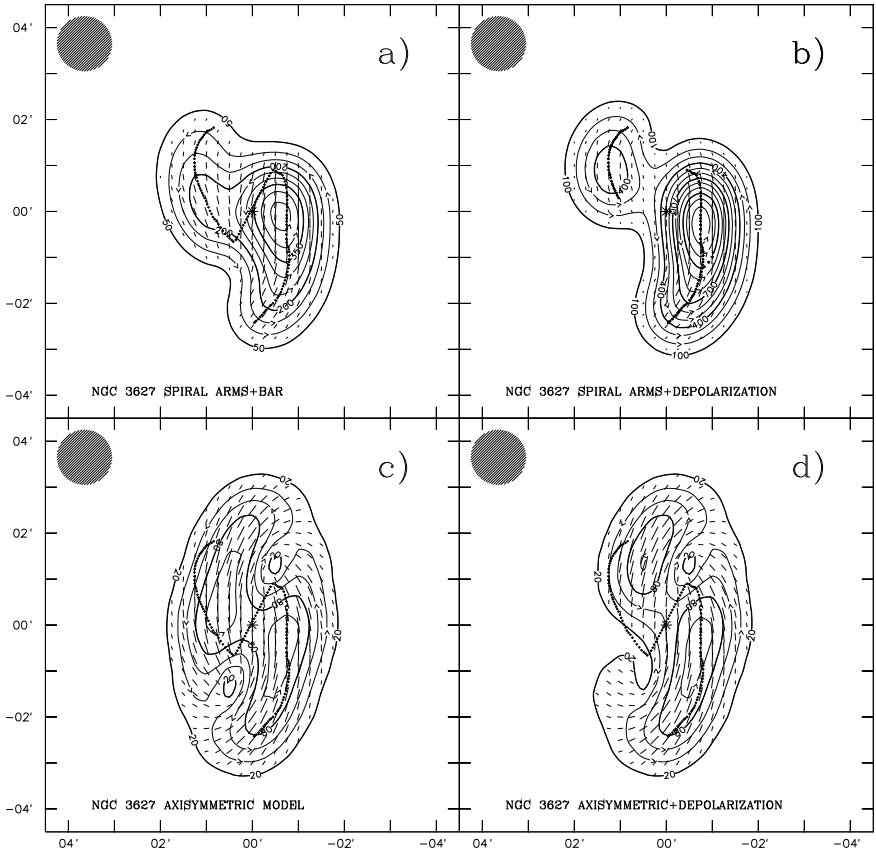}} 
\parbox[b]{55mm}{ 
\caption{ 
The model distributions of polarized intensity in NGC~3627 assuming: 
a)~a regular magnetic field parallel to the arms and concentrated in 
dust lanes, as well as one parallel to the bar, b) as above, without the 
contribution from the bar and no regular field in the strongly 
star-forming segment of the eastern arm, c)~a disk-parallel axisymmetric 
field with a constant pitch angle of $30\degr$, d)~the above 
axisymmetric field with no polarization at azimuthal angles 
corresponding to the star-forming arm segment. For spiral arm models 
(a~and b) the thick dotted lines mark the parts of spiral structure 
assumed to house regular fields, for axisymmetric models (c~and d) the 
spiral pattern is also shown for easier comparison with the 
observations. The models were made for $r_0=1\farcm 2$, $p_{1}=0$ and 
$p_{2}=0.5$. 
} 
\label{fig9} }
\end{figure*} 

The eastern and western polarized lobes differ not only in their peak 
brightness but also in their positions and azimuthal extent relative to 
optical arms. Moving in the galactic disk along the azimuth 
anticlockwise from the corresponding bar ends, we observe initially a 
very similar increase of polarized brightness (Fig.~7). However, while 
the polarized intensity in the western arm continues to rise reaching a 
maximum at an azimuthal distance of about $100\degr$ from the southern 
bar end, the polarized brightness in the eastern arm drops at an 
azimuthal distance of $75\degr$ from the northern bar end, showing even 
a local minimum at $100\degr$. As the inner parts of both arms have a 
similar shape the unpolarized region in the eastern arm does not result 
from effects related to the spiral arm geometry. The statistical 
significance of the differences between the profiles was estimated by 
averaging them in non-overlapping azimuthal angle intervals 
corresponding to the beam size at the appropriate azimuthal distance 
from the bar end. This yielded for each profile 7 statistically 
independent points. Assuming that they represent independent random 
variables having the r.m.s. dispersion equal to the polarization map 
noise we found that the probability that the differences between 
profiles result purely from random fluctuations is smaller than 
$2\times 10^{-6}$. This result was checked to be independent of the starting 
point of averaging intervals. 

The dust lanes in the western arm and its segment in the inner part of 
the eastern one coincide with ridges of CO(1-0) emission (Reuter et~al. 
1996, see also Fig.~8a) tracing very dense, narrow and elongated 
molecular gas complexes forming in density-wave compression regions. 
They are also visible in the HI map of Zhang et~al. (1993). While the 
western polarized lobe peaks on the CO ridge and extends along it, the 
eastern one falls on a hole in the CO emission. On the other hand, the 
CO (and HI) ridge in the southern part of the eastern arm, being even 
stronger than the western one, coincides with a completely unpolarized 
region. The narrow, elongated CO features are thus not always associated 
with highly polarized regions, as one would expect from pure compression 
of magnetic field by density waves. We note however, that the western CO 
ridge is accompanied by only isolated, small HII regions (Smith et~al. 
1994, Fig.~8b) while the unpolarized CO ridge in the eastern arm hosts a 
chain of large complexes of H$\alpha$-emitting gas. Faraday effects at 
10.55~GHz are negligible, thus the depolarization is primarily of 
geometrical nature. Tangling of the magnetic field by star-forming 
process inside the H$\alpha$-bright knot is insufficient: most of the 
star formation occurs outside of the dust lane, thus it cannot destroy a 
possible density wave-related field and occupies a too small volume to 
randomize a smoothly-distributed dynamo field in a whole disk quadrant. 
However, at the inclination of $67\fdg 5$ the polarization degree may 
be significantly lowered by vertical magnetic field fluctuations 
developing above the strongly star-forming chain in the eastern arm. 
They may be caused by vertical chimneys or superbubbles powered by 
multiple supernova explosions (see e.g. Mineshige et al. 1993, Tomisaka 
1998). Some role of Parker instabilities (Parker 1966) cannot be 
excluded, too. The magnetic field structures stretching perpendicularly 
to the disk with a significant vertical field component, projected to 
the sky plane and seen by a large beam together with the disk-parallel 
field (either concentrated in the dust lane or filling the whole disk), 
may provide an efficient depolarizing agent. 

\subsection{Magnetic field models} 

To judge whether our polarization map is dominated by the density-wave 
magnetic field component or by the axisymmetric, dynamo-type we need the 
beam-smoothed models of polarized emission from magnetic fields of an 
assumed structure. Four kinds of models were computed using techniques 
described by Urbanik et~al. (1997): 

\begin{itemize} 
\item[a)] A model assuming a regular magnetic field concentrated in 
 prominent dust lanes and running along them. In addition the 
 magnetic field running along the bar could be switched on and off. 
\item[b)] The above model without polarized emission from the strongly 
 star-forming segment of the eastern arm. 
\item[c)] A model assuming an axisymmetric, spiral, plane-pa\-ra\-llel 
 field with a constant intrinsic pitch angle of $-30\degr$ (mean 
 value for NGC~3627). 
\item[d)] The above model without polarized emission in the eastern 
 region corresponding to the discussed star-for\-ming arm segment, as 
 expected for strong vertical field fluctuations seen in projection 
 together with a disk-para\-llel field. 
\end{itemize} 

In all models the adopted radial distribution of the total field 
strength and cosmic ray electron density was set to yield an exponential 
total power disk with a radial scale length $r_0$ (Sect.~4.1). The 
intrinsic degree of polarization was rising linearly from $p_1$ in the 
centre to $p_2$ in the disk outskirts. $r_0$, $p_1$~and $p_2$ were 
adjusted to yield the best qualitative agreement of the models with 
observations. 

The best results presented in Fig.~9a--d are as follows: 

\begin{itemize} 
\item[-] The presence of two polarized lobes with B-vectors running 
parallel to optical spiral arms is reproduced by both axisymmetric and 
spiral arm models. They both give the position of the western lobe on 
the middle of the spiral arm, in agreement with observations. 
\item[-] Both axisymmetric and spiral arm models need an extra 
depolarizing agent in the star-forming segment of the eastern arm, 
otherwise both models give the maximum of polarized intensity where 
observations show a complete lack of polarization (Fig.~9a and c). 
\item[-] Only the axisymmetric models (with and without an extra 
depolarization, Fig.~9c and d) correctly place the NE lobe in the 
interarm space. The spiral arm models (Fig.~9a and b) invariably give 
the position of the eastern polarized lobe on the position of the 
prominent dust lane which disagrees with observations. 
\item[-] Only the axisymmetric models reproduce the observed regions of 
 a weak polarized signal encircling minima at the bar ends with 
B-vectors turning smoothly from one arm to the other. 
\item[-] Even with a suppressed polarization in the SE disk region the 
axisymmetric model yields similar peak amplitudes of both lobes and 
thus does not reproduce their observed asymmetry. The spiral arm model 
does this considerably better. 
\item[-] Another difficulty of the axisymmetric model is a too large 
extent of the modelled lobes into the outer disk compared to their 
rather peaked shape in NGC~3627 (especially of the western one). 
Varying $r_0$ and/or $p_2$ can make the western lobes more peaked but 
moves it to the interarm space, worsening the agreement with 
observations. 
\end{itemize} 

The last two difficulties of the axisymmetric model can be somewhat 
relaxed by adding an unresolved polarized source in the middle of the 
western arm, where its unusually straight part (see Fig.~5) and a steep 
HI gradient on this disk side are suggestive for an external gas 
compression (Haynes et~al. 1989). However, the difficulties of the 
spiral arm model can only be improved by adding a widespread, 
significant axisymmetric magnetic field. 

\begin{figure} 
\resizebox{\hsize}{!}{\includegraphics{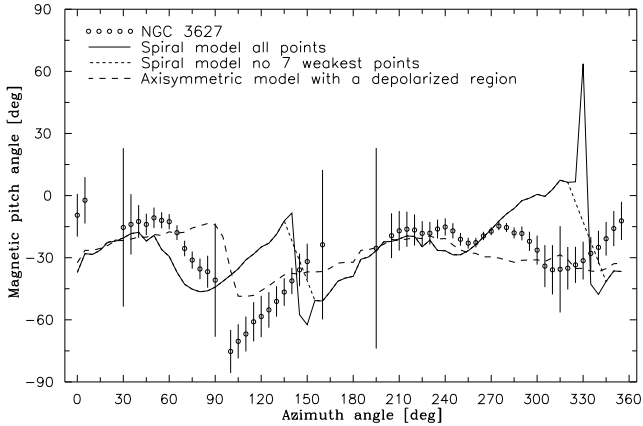}} 
\caption{ 
The azimuthal variations of the B-vector pitch angles $\psi$ corrected 
to face-on position, averaged in azimuthal sectors $5\degr$ wide along a 
ring with a width of 24\arcsec and a face-on radius of 2\arcmin. The 
data for NGC~3627 are shown by circles and bars, for the spiral arm 
model by a solid line and for the axisymmetric model with an unpolarized 
region in the eastern arm segment as a dashed line. A dotted line shows 
an attempt to remove the jumps of $\psi$ in the spiral model by dropping 
sectors of lowest signal from the analysis. For NGC~3627 only the data 
above the $1\sigma$ noise level were used to determine the pitch angle. 
} 
\label{fig10} 
\end{figure} 

Attempts to reproduce the variations of face-on corrected magnetic pitch 
angles $\psi$ with the azimuthal angle in the disk are shown in Fig.~10. 
The observed changes of $\psi$, and especially a jump near the azimuthal 
angle of $90\degr$, rule out a purely axisymmetric magnetic field. 
However, addition of the discussed unpolarized region to our 
axisymmetric model yields the jump at the correct position, though its 
exact shape in our simple model is still far from reality. The spiral 
arm models shown in Fig.~9a~and b also have some dip at about 90$^{o}$, 
however they yield abrupt jumps of $\psi$ at $150\degr$~and $330\degr$. 
These features do not depend on model parameters, nor on the inclusion 
or exclusion of low-brightness regions in model maps. They naturally 
result from the spiral arm shape and cannot be removed without changing 
the basic model assumptions. In the azimuthal angle range of 
$270\degr$~to $330\degr$ the spiral arm model deviates also from the 
data much more than that assuming the axisymmetric field. 

Despite very strong density waves NGC~3627 still shows clear signatures 
of axisymmetric, dynamo-type magnetic fields. At present it is hard to 
say whether it dominates the disk field, showing only locally effects of 
external compression, or whether it coexists with the density-wave 
component as an important constituent of the global magnetic field. A 
detailed discrimination between these possibilities needs observations 
with a considerably higher resolution complemented by extensive 
computations of a whole grid of detailed quantitative models of 
NGC~3627, which is beyond the scope of this paper. 

\section{Summary and conclusions} 

The strongly interacting Leo Triplet galaxy NGC~3627 has been observed 
at 10.55~GHz with the 100-m MPIfR radio telescope. Total power and 
polarization maps with a resolution of 1\farcm 13, very sensitive to 
extended, diffuse polarized emission were obtained. Their analysis in 
the context of optical, CO and H$\alpha$ data yielded the following 
results: 

\begin{itemize} 

\item[-] The total power map shows two bright maxima at the bar ends, 
coincident with strong CO and H$\alpha$ peaks. There is no evidence of a 
significant radio emission from the central region, thus a large central 
molecular complex (Reuter at~al. 1996) has star formation rate much 
lower than the molecular gas accumulations at the bar ends and its weak 
H$\alpha$ emission is not entirely due to a strong dust obscuration. 
However, differences in absorption could explain the asymmetry of 
H$\alpha$ emission between the bar ends. 

\item[-] The polarized emission forms two asymmetric lobes: a strong one 
peaking on the dust lane bent inwards in the western arm and extending 
along this arm while a weaker one is located in the interarm space in 
the NE disk. We also detected diffuse, extended, polarized emission 
encircling the bar ends away from spiral arms. The polarization 
B-vectors run parallel to the principal arms, twisting around the bar 
ends in weakly polarized regions. 

\item[-] The southern part of the eastern arm is completely depolarized. 
This region shows signs of strong density wave compression however, it 
contains a lot of ionized gas indicating strong star formation. At the 
galaxy's inclination, the development of vertical magnetic instabilities 
seen in projection together with the disk-parallel magnetic field could 
be a suitable depolarizing agent. 

\item[-] Attempts to qualitatively explain the distribution of polarized 
intensity and the B-vector geometry in NGC 3627 in terms of simple 
magnetic field models suggest the presence of a significant (if not 
dominant) axisymmetric, dynamo-type field. However, to best explain our 
polarization maps all models need an extra geometrical depolarization 
e.g. by vertical fields above the discussed star-forming segment of the 
eastern arm. To reproduce the polarization asymmetry the 
dynamo-generated field also requires an extra polarized component 
(probably due to external compression?) at the position of the straight 
western arm segment. 

\end{itemize} 

The present work demonstrates that even in galaxies with strong density 
waves observations sensitive to extended diffuse polarized emission 
cannot be fully explained by the density wave-related magnetic field 
component but show clear signatures of large-scale axisymmetric 
dynamo-type fields. On the other hand, in the same object the 
density-wave component may show up much better or become dominant in 
high-resolution interferometric observations underestimating the 
extended polarized emission. We believe that combined interferometric 
and single-dish data on such objects supported by extensive modelling 
might help to establish the mutual relationships and relative roles of 
turbulence and density-wave flows in galactic magnetic field evolution. 

\begin{acknowledgements} 
The Authors wish to express their thanks to Dr~Beverly Smith from IPAC 
for providing us with her H$\alpha$~map in a numerical format. We are 
grateful to numerous colleagues from the Max-Planck-Institut f\"ur 
Radioastronomie (MPIfR) in Bonn, in particular to Drs E.M.~Berkhuijsen 
and P.~Reich for their valuable discussions during this work. M.S. and 
M.U. are indebted to the Directors of the MPIfR for the invitations to 
stay at this Institute, where substantial parts of this work were done, 
and to Dr H-P.~Reuter for his assistance in using his CO maps. They are 
also grateful to colleagues from the Astronomical Observatory of the 
Jagiellonian University in Krak\'ow and in particular to Drs. 
K.~Otmianowska-Mazur and M.~Ostrowski for their comments. We thank to 
the anonymous referee for the valuable remarks. This work was supported 
by a grant from the Polish Research Committee (KBN), grants no. 
578/P03/95/09. and 962/P03/97/12. Large parts of computations were made 
using the HP715 workstation at the Astronomical Observatory in Krak\'ow, 
partly sponsored by the ESO C\&EE grant A-01-116 and on the Convex-SPP 
machine at the Academic Computer Centre ''Cyfronet`` in Krak\'ow (grant 
no. KBN/C3840/UJ/011/1996 and KBN/SPP/UJ/011/1996). 
\end{acknowledgements}

\end{document}